# COGNITIVE RADIO RESOURCE SCHEDULING USING MULTI-AGENT Q-LEARNING FOR LTE

Najem N. Sirhan and Manel Martinez-Ramon

Electrical and Computer Engineering Department, University of New Mexico, Albuquerque, New Mexico, USA

## ABSTRACT

*In this paper, we propose, implement, and test two novel downlink LTE scheduling algorithms. The implementation and testing of these algorithms were in Matlab, and they are based on the use of Reinforcement Learning (RL), more specifically, the Q-learning technique for scheduling two types of users. The first algorithm is called a Collaborative scheduling algorithm, and the second algorithm is called a Competitive scheduling algorithm. The first type of the scheduled users is the Primary Users (PUs), and they are the licensed subscribers that pay for their service. The second type of the scheduled users is the Secondary Users (SUs), and they could be un-licensed subscribers that don't pay for their service, device-to-device communications, or sensors. Each user whether it's a primary or secondary is considered as an agent. In the Collaborative scheduling algorithm, the primary user agents will collaborate in order to make a joint scheduling decision about allocating the resource blocks to each one of them, then the secondary user agents will compete among themselves to use the remaining resource blocks. In the Competitive scheduling algorithm, the primary user agents will compete among themselves over the available resources, then the secondary user agents will compete among themselves over the remaining resources. Experimental results show that both scheduling algorithms converged to almost 90% utilization of the spectrum, and provided fair shares of the spectrum among users.*

## KEYWORDS

## 1. INTRODUCTION

### 1.1. Research Problem

According to the Federal Communication Commission (FCC) in [8], when mobile operators use the fixed spectrum assignment of spectrum resources, these spectrum resources may be under-utilized as shown in Figure 1. This is because of the irregular demands by licensed subscribers that vary according to time and geographical area. In other words, Spectrum underutilization can occur when some resource blocks that are assigned to licensed users at some particular times are not being used. These un-used resource blocks are called white spaces or spectrum holes. To solve this underutilization problem and make the most out of a spectrum, Dynamic Spectrum Sharing (DSS) that is based on dynamic spectrum assignment has to be deployed instead of the fixed spectrum assignment. This made a lot of scientists research on the different implications of communication and signal processing that are needed for Dynamic Spectrum Access (DSA) networks. DSA is a set of techniques that aims to better utilize the use of the licensed spectrum by detecting the spectrum holes due to underutilizing the use of it and allowing unlicensed users to use it as well [1] [28]. The concepts of spectrum holes and dynamic spectrum access are shown in Figure 2.

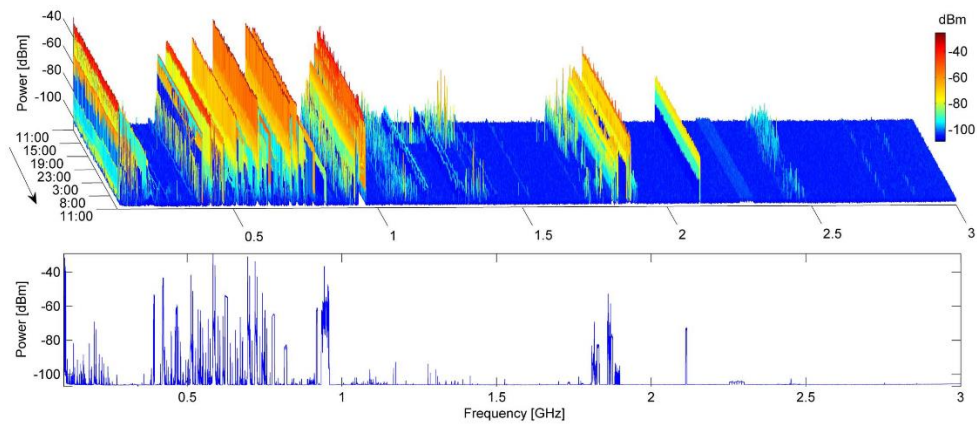

Figure 1. Spectrum occupancy in the suburb of Brno as measured by a team of researchers for a six-day period. In the upper graph, the maximum power obtained during this measurement period is displayed. In the lower chart, the power profile for the average power of the upper chart is displayed [23].

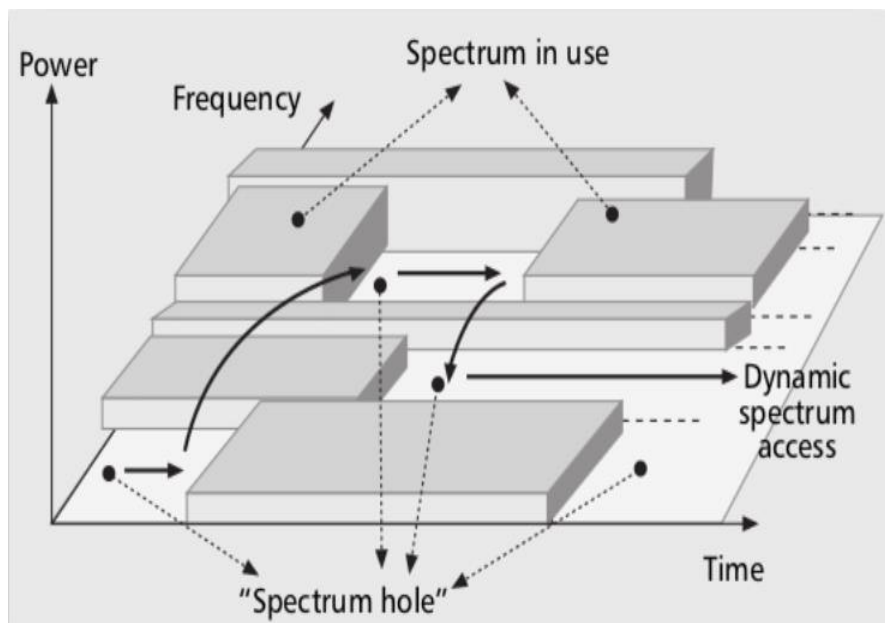

Figure 2. The concept of spectrum holes is used to refer to an unused resource of the spectrum, and how the use of dynamic spectrum access can be used to allow other type of users to utilize the use of these resources [2]

The research problem which this paper proposes two algorithms to solve can be formulated as follows; scheduling two types of users over the downlink radio channels, the primary users over the available radio resource blocks, and the secondary users over the remaining resources. The scheduling process is done every Transmission Time Interval (TTI), which equals 1 ms. Each user whether it's a primary or secondary is considered an agent. The primary user agents will use the collaborative approach in order to make a joint decision about allocating the resource blocks to them, or the competitive approach depending on the scheduling algorithm that is used. And the remaining resource blocks will be sensed by the secondary user agents, and then to be used. The secondary user agents will use the competitive approach to compete among themselves to use these remaining resource blocks. Both the primary and the secondary users can access the network pool of radio resources, that is provided by the macro-cell. However, the secondary users have lower priority than the primary users, and they are transparent to primary users, which means that the primary users will consider that the whole resource blocks are available to them only, and they will access these resources as if the secondary users don't exist. And in the cases if a secondary user is using an available radio resource at a current time slot, but a primary user is about to use the same radio resource at the next time slot, the secondary user has to withdraw and free this resource to be used by the primary user.

## 1.2. Cognitive Radio (CR)

According to [13], Cognitive Radio (CR) is defined as a form of wireless communication in which a transceiver can intelligently detect which communication channels are in use and which are not, and can be dynamically configured to move into a vacant channel while avoiding occupied ones.

Cognitive radio systems were designed to optimally use the electromagnetic spectrum through their ability to detect vacant channels and move into them. This goal is achieved through a learning cycle that cognitive radio systems go through. This learning cycle consists of three main stages, perception, learning, and reasoning, and it is shown in Figure 3. Perception is the first stage of the learning cycle, and it starts by sensing the spectrum in order to collect data about the surrounding radio environment, e.g., the channels' conditions and their availabilities. CR systems should not only sense and are aware of the medium, but it should also have the ability to learn and reason. Learning is the second stage and most important stage, and this is because it includes transforming the obtained information about the radio environment into knowledge through the use of classification methodologies. The reasoning is the final stage in this cycle, in which the obtained knowledge is used to make decisions that meet with the cognitive radio objectives, e.g., optimizing the usage of the spectrum to maximize the system's throughput [13] [22].

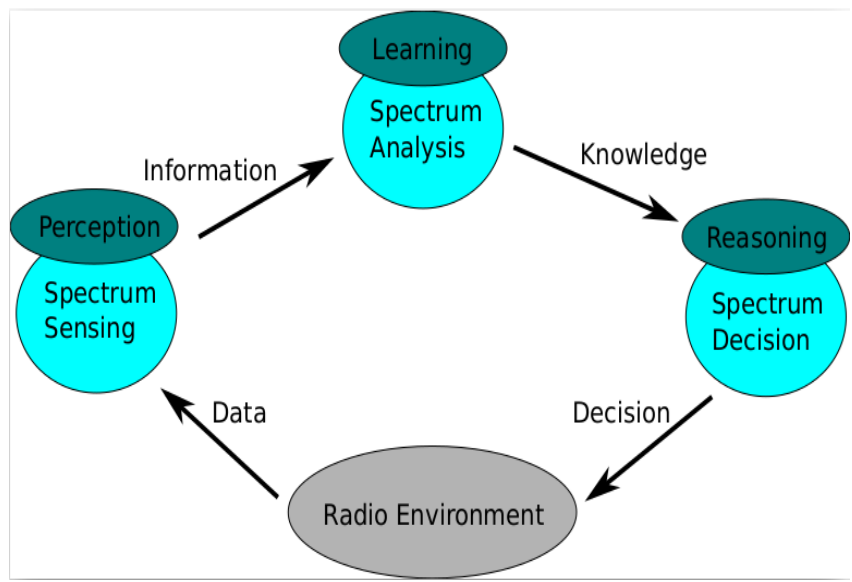

Figure 3. Cognitive radio learning cycle

## 1.3. Reinforcement Learning (RL)

Reinforcement Learning (RL) could be explained in its most simplistic terms, it implies a situation in which a learning agent is left alone and not told what to do, and it has to form a basic understanding of its environment. RL offers a powerful set of tools for sequential decision-making under uncertainty. In RL, there are five main elements, the learning agents, actions, policy rules, rewards, and the system model. The learning agent learns what to do "what action to perform" in a situation "state" in an environment by trying an action despite being uncertain about the action's effect on the environment in that situation. Then, the learner maps the effect of this action in that situation for the environment with the performed action in order to maximize a numerical reward signal. This mapping between the actions and rewards is called the policy rules which define the behavior of the learning agent [22] [14].

An agent was defined by [18] as autonomous to the extent that its behavior is determined by its own experience. A more precise definition of a learning agent is found in [9], in which a learning agent is an agent that develops through own experience and with the help of some built-in knowledge, in which an action policy will be directly mapped from its observations and internal conditions.

Now, as regards the use of RL in CR, it is recommended for spectrum sensing and Medium Access Control (MAC) protocols, as in [27]. RL plays a key part in the development of cognitive radio systems, and it forms the framework upon which cognitive radio systems are built. This is because the basic and most important characteristic of cognitive learning is its ability to learn in an autonomous way [5].

## 1.4. Q-Learning and its use in CR

Q-learning which was proposed by [25] is a model free reinforcement learning techniques, it could be used even if the channels were not known or the channels Markov model was not known, which means that the learning agent doesn't need to know the Markov chain of all the previous states of the channels in-order to take an action, it uses what is called a Q-table instead. In Q-learning based system models, learning agent exists in an environment which have a finite number of states, which keeps changing according to the actions that are taken by the agent or other existing agents in the same environment. These different states form the set of states $S$. The learning agent can sense it's current state $s_t$ at time $t$, in which $s_t \subset S$. Based on this current state, the learning agent will choose an action $a$ from a set of actions $A$ to be executed in time $t+1$. Based on this executed action and its effect on the environment, a reward function $r_{t+1}(s_t,a)$ will be calculated, the higher the reward the higher the probability of choosing this performed action [22] [12] [15].

In each Q-learning-based model, a Q-table $Q(s,a)$ is constructed and updated. The formula that is used to update an entry in the Q-table is called the Q-Learning formula as proposed by [25], and it is as follows:

$$Q(s_t, a_n) \leftarrow (1-\alpha)\, Q(s_t, a_n) + \alpha[(r_{t+1}\,(s_t, a_n) + \gamma \max_{ai} Q(s_{t+1}, a_i)] \qquad (1)$$

In cognitive radio, the environment is the radio channels and their states, $s_t \subset S$, $s_t$ will represent the availability of these channels at time $t$, and $S$ will be the set of all the radio channels' states at all time. Each radio channel will have two states, either Idle ($I$), or Busy ($B$).

The action $a$ of selecting radio channels belong to the actions set $A$, $a \subset A = \{a_1, ..., a_n\}$. The decision policy of choosing an action $a$ in time period $t$, to be executed in time period $t+1$ is based on a maximum argument, $\max_{ai} Q(s_{t+1}, a_i)$, where $i = \{1, ..., n\}$, $a_i$ representing all the actions in the action set at state $s_{t+1}$.

The reward $r_{t+1}$, is the immediate reward that is obtained after executing action $a$ in state $s_{t+1}$. $r_{t+1} \subset r$, where $r$ is the set of rewards. $r_t$ is defined by the system designer, it could be defined as the actual achieved user's communication throughput, or it could be defined as the Jain's fairness index.

The learning rate is denoted by $\alpha \subset (0,1)$. The discount factor is denoted by $\gamma \subset [0,1)$, and it determines the importance of future reward.

## 1.5. Multi-Agent Q-Learning and its use in CR

In multi-agent Q-learning algorithms, learning agents either collaborate or compete among each other. In the case of collaborative learning agents, the system is called a collaborative multi-agent system, and in the case of competitive learning agents, the system is called a competitive multi-agent system. The main difference between these two systems is that the collaborative learning agents' decision making policies are based on maximizing the joint reward. While, the competitive learning agent's decision making policies are based on maximizing their own individual rewards [17].

In the case of the collaborative approach, part of the channel has to be reserved for information sharing. However, in the case of competitive approach, there is no need for information sharing. In addition to this, in the collaborative approach, there is one centralised node that contains one Q-table for all the agents, in which they cooperate by sharing their sensing information in order to update the centralised shared policies. However, in the competitive approach, each agent senses the medium by considering other agents part of the Radio Frequency (RF) environment, and each agent updates its own Q-table [22] [3] [4].

## 2. RELATED WORK

In the literature, there exist many packet scheduling algorithms that are based on calculating a utility function. Each of these algorithms has its own unique utility function, whether to increase the throughput, or to decrease the delay, or to improve the fairness. The utility function calculation is the core process of scheduling with this type of algorithms. This is because it contains all the necessary parameters which are included in calculating the best scheduling decision that meets with the radio environment states and the users' requirements [19][20][21].

In addition to packet schedulers that are based on calculating a utility function, there are other types of packet schedulers found in the literature that are based on the use of the Q-learning algorithm. However, they exist in a much lesser amount. In Q-learning-based LTE scheduling algorithms, constructing a Q-table is the core process of scheduling with this type of algorithms. Constructing and updating the Q-table depends on three main elements, the state, action, and reward. The reward is calculated after executing an action by a learning agent in a certain time and the radio environment state. Assigning the reward function is crucial in the agent's learning process and in regulating its behaviour. This is because calculating the reward indicates to the learning agent how much gain it can get by executing an action and how much effect it has on the environment. The repeated cycle of the agent's learning process will update and construct the Q-table, in which it will contain optimal action policies that will eventually lead to an optimized scheduling decision.

In [6], the authors proposed an algorithm that uses RL technique to choose a scheduling rule from a pool of scheduling rules, in which the action was defined as what scheduling algorithm to use, and the reward was calculated based on the system throughput, system capacity, and spectral efficiency. The authors modelled their work theoretically without implementing the model in a simulation environment.

In [7], the authors have proposed a scheduling algorithm that is based on using the Q-learning approach. They use Q-learning to choose which scheduling algorithm to use for scheduling the resources among users. The authors aimed at their scheduling algorithm to achieve a trade-off between throughput and fairness. In their model, their Q-table's entries consist of two elements, the action and the obtained reward. Their action is what scheduling algorithm to choose, and the reward is based on calculating the average throughput and fairness. They used the LTE-Sim [16] simulator in order to do their simulation. Their results were in terms of the average normalized system throughput.

In [11], the authors proposed a scheduling algorithm that is based on the use of Q-learning, and they implemented, and run experiments to test it using the LTE System level Simulator [10]. The authors proposed two forms of their algorithm. One for a single agent Q-learning platform, in which the eNodeB acts as an agent. And another one for a multi-agent Q-learning platform, in which each eNodeB is an agent, and all these agents coordinate with each other in a harmonized way, and that is why they called their algorithm, the Harmonized Q-learning algorithm. Their results were in terms of user wideband Signal to Noise Ratio (SINR), and the average user spectral Efficiency.

In [24], the authors presented two Q-learning approaches in allocating the unused radio resources by the licensed users to unlicensed users. The two Q-learning approaches were the cooperative Q-learning and non-cooperative Q-learning approach, and they presented these approaches in the form of formulas. The authors also evaluated these approaches by running tests over their model. In their model, they did not use licensed users, but they mimicked their presence by varying the number of unused radio resources. In their evaluation tests, they included four access schemes for unlicensed users to access the unused radio resources. The four access schemes were, the random, non-cooperative, partial cooperation, in which some users cooperate and others don't cooperate, and full cooperative access scheme. Their evaluation test results showed that the full cooperation access scheme had the best results in terms of throughput and fairness, then followed by the partial cooperation, then followed by the non-cooperative, and finally followed by the random access scheme which had the lowest results.

## 3. PROPOSED SOLUTION 1 – COLLABORATIVE SCHEDULING ALGORITHM

The first scheduling algorithm that is proposed and implemented in Matlab is the Collaborative scheduling algorithm, Algorithm 1. In this algorithm the scheduling process is divided into two stages; the first scheduling stage is performed for the primary users, and the second one, which comes after is performed for the secondary users. It combines two Q-learning approaches, the collaborative approach that is based on modifying the work proposed by [5] for primary users, and the competitive approach that is based on the work proposed by [25] for secondary users.

---

**Algorithm 1** Collaborative Scheduling Algorithm

---

**Input**: no. PUs, no. SUs, no. RBs, no. epoch, exploration parameter $e$
**Output**: PUs shared Q-table, SUs Q-tables
Initialize all parameters: Q-tables, PU Actions set $A_{PU}$, SU Actions set $A_{SU}$, state of RBs set $S$

**for** $t = 1$ to no. epoch **do**

**for** $i$ = 1 to N, where N = no. PUs **do**

    **if** no. available RBs > 0 **then**

        **if** rand < = $e$ **then**

            PU agent $i$ will explore by taking an action $a^i$ randomly from $A_{PU}$

        **else**

            PU agent $i$ will exploit by taking an action $a^i$ from $A_{PU}$ based on the Expectation Values table

        **end if**

        $A_{-i} = A_{PU} - \{a^i\}$

    **end if**

    Mark the used RBs in the sub-state $s_t$ as busy for PU agent $i$

**end for**

Form the joint action ($a^{-i} \cup a^i$) to be executed in the following state

Calculate the obtained reward $r_{t+1}$ after excuting the joint action ($a^{-i} \cup a^i$) according to

the Jain's fairness index: $r_{t+1} = ( ( \sum_{i=1}^{N} T_i )^2 ) / ( N \sum_{i=1}^{N} T_i^2 )$

$T_i$ denotes the throughput obtained for primary user $i$. N is the number of primary users which is 5

Update the shared Q-table according to the following Q-learning formula:

$Q(s_t, (a^{-i} \cup a^i )) \leftarrow (1 - \alpha) \ Q(s_t, (a^{-i} \cup a^i )) + \alpha[(r_{t+1} (s_t, (a^{-i} \cup a^i )) + \gamma \ V(s_{t+1})]$

**for** $i$ = 1 to N **do**

    PU agent $i$ will update its counters about other PU agents taking their actions

    PU agent I will calculate the product of probabilities of other PU agents taking their actions: $\prod j \neq i \{Pr^i_{a^{-i}[j]}\}$

    PU agent $i$ will update the Expectation Value of its indivisual action $a^i$ in its expectation value table:

$$EV(a^i) = \sum_{a^{-i} \in A_{-i}} Q(a^{-i} \cup a^i) \prod j \neq i \{Pr^i_{a^{-i}[j]}\}$$

**end for**

**for** $k$ = 1 to no. SUs **do**

    **if** no. free RBs in the sub-state $s_t$ > 0 **then**

        **if** rand < = $e$ **then**

            SU agent $k$ will explore by taking a random action $a^k$ from $A_{SU}$

        **else**

            SU agent $k$ will exploit by taking a greedy action $a^k$ from $A_{SU}$

        **end if**

        SU agent $k$ will calculate the reward of executing action $a^k$

        SU agent $k$ will update its Q-table based on the Q-Learning formula:

        $Q(s_t, a^k) \leftarrow (1 - \alpha) \ Q(s_t, a^k) + \alpha[(r_{t+1} (s_t, a^k) + \gamma \ V(s_{t+1})]$

    **end if**

    $A_{-k} = A_{SU} - \{a^k\}$

**end for**

**end for**

## State set S

The set $S$ represent all the observed states of all the Resource Blocks (RBs) at all the running time.

$S = \{s_0, s_1, ..., s_{t-1}, s_t, s_{t+1}, ..., s_{no.epoch} \}$

$s_t = \{RB1, ..., RBx\}$, where $x = no.RBs$

$s_t \subset S$, $s_t$ is a sub-set of the set $S$ that represent the states of all the available RBs at time $t$. And it is the same for all users agents, the primary users agents, and the secondary users agents. Each resource block has three states, free, or busy for a primary user, or busy for a secondary user.

**Primary users scheduling stage**

The primary users scheduling stage starts by each primary user agent taking an action from the primary users action set, then these actions will form one joint action. Then the scheduler will calculate the obtained reward from executing this joint action. Then the scheduler will update the PUs shared Q-table.

**Actions and Actions set**

The primary user actions set $A_{PU}$ consists of multiple actions, each action has a different number of resource blocks with unique indexes. This will avoid collisions between primary users agents since they all have access to this action set.

$A_{PU} = \{a_1{}^{PU}, ..., a_n{}^{PU}\}$, $a^{PU}$ is an action to be taken by primary user agent, $n$ is the number of actions in $A_{PU}$.

In regard to the decision policy that is deployed for choosing an action, it is based on either exploration or exploitation, and this is determined based on the value of the exploration parameter $e$. The value of this parameter determines the probability of exploration and the probability of exploitation. For example, if $e = 0.5$, then there is a 50% probability of an agent to explore, and 50% probability of an agent to exploit.

In the case of exploration, the agent will make a random choice in taking an action from the primary user action set. In the case of exploitation, the agent will make its choice to take an action based on the expectation formula values in the joint Q-table, to execute an individual action which forms with other agents' actions the best joint action that exists in the shared Q-table. The best joint action will be associated with the highest reward.

After the first primary user's agent takes an action, the scheduler will update the primary users action set as, $A_{-i} = A_{PU} - \{a^i\}$, so that each agent will take a different action, this will help in avoiding any collision between primary users. Then the next primary user agent will enter the same loop, and this loop will be repeated until all primary user agents take action. The result of all the individual actions that were taken by the five primary user agents will contribute in forming one joint action, to which we denote it by $(a^{-i} \cup a^i)$.

**Rewards and Rewards set**

The reward set $R_{PU}$ consists of all the obtained rewards at all the running time.

$R_{PU} = \{r_1, ..., r_{t-1}, r_t, r_{t+1}, ..., r_{no.epoch}\}$. $r_{t+1}$ is the reward that is obtained after executing the joint action $(a^{-i} \cup a^i)$ in state $s_{t+1}$.

The reward function for the primary user agents is chosen to optimize the primary user's average throughput for all the users while at the same time maintaining a fair share of the radio resources to each user. So in this work, the scheduler calculates the obtained reward $r_{t+1}$ according to Jain's fairness index:

$$r_{t+1} = ( \ ( \ \textstyle\sum_{i=1}^{N} \quad T_i \ )^2 \ ) \ / \ N \ \textstyle\sum_{i=1}^{N} \quad T_i^2 \ ) \tag{2}$$

Jain's fairness index, will rate the fairness of distributing the throughput "denoted by $T_i$" among primary users, where N is the number of primary users.

## Q-table and Expectation values

After executing the joint action and calculating the obtained reward, the scheduler will update the shared Q-table. The Q-learning formula that is used to update an entry in the Q-table is as follows:

$$Q(s_t, (a^{-i} \cup a^i)) \leftarrow (1 - \alpha) \ Q(s_t, (a^{-i} \cup a^i)) + \alpha[(r_{t+1} (s_t, (a^{-i} \cup a^i)) + \gamma \ V(s_{t+1})] \tag{3}$$

Where $V(s_{t+1})$ is determined by the policy of choosing an action at time $t$, and $a^i$ represent the action $a$ taken by primary user $i$, $a^{-i}$ represent the actions taken by all primary users other than primary user $i$, $(a^{-i} \cup a^i)$ represent the joint action of all the actions taken by the primary users. After the scheduler updates the shared Q-table, each PU agent will do the following:

<u>First</u>, it will update its counters about other PU agents taking their actions. The primary user agent's is a Joint Action Learner (JAL), which means that the agent learns about other agents actions and their effect, in addition to the effect of its action. In order for the agent $i$ to learn about other agents actions and their effects, it keeps a count $C^{-i}{}_{a^{-i}}$ for the number of times other agents "which we denote to any one of them by $-i$" has taken action $a$ in the past. Then agent $i$ calculates the probability of agent $-i$ to take an action $a^{-i}$ as the following formula:

$$Pr^i{}_{a^{-i}} = C^{-i}{}_{a^{-i}} / \ ( \ \textstyle\sum_{b^{-i} \in A_{-i}} \quad C^{-i}{}_{b^{-i}} \ ) \tag{4}$$

Where $b^{-i}$ represent all the previous actions taken by agent $-i$.
<u>Second</u>, it will calculate the product of probabilities of other PU agents taking their actions as the following formula:

$$\textstyle\prod j \neq i \{ Pr^i{}_{a^{-i}[j]} \} \tag{5}$$

<u>Third</u>, it will update the expectation value $EV(a^i)$ of its individual action $a^i$ that it took in its Expectation Values table as the following formula:

$$EV(a^i) = \textstyle\sum_{a^{-i} \in A_{-i}} \quad Q(a^{-i} \cup a^i) \prod j \neq i \{ \ Pr^i{}_{a^{-i}[j]} \} \tag{6}$$

These expectation values help the PU agent in implementing its exploitation strategy.

## Secondary users scheduling stage

The secondary users scheduling stage does not include any cooperation between the secondary user agents. On the contrary, the secondary user's agents compete on the remaining resource blocks that are left after scheduling the primary users. The secondary users scheduling loop of taking an action, then obtaining a reward based on executing this action, and then updating the Q-table, is repeated for every secondary user agent.

**Actions and Actions set**

The secondary user actions set $A_{SU}$ is different from the primary user actions set. It is accessed by all the secondary user agents. It consists of multiple actions, and each action has a different number of resource blocks in order to avoid collision between secondary users.

$A_{SU} = \{a_1{}^{SU}, ..., a_m{}^{SU}\}$, $a^{SU}$ is an action to be taken by a secondary user agent, $m$ is the number of actions in $A_{SU}$.

These actions are designed with an upper limit on how many resource blocks the agent can get. This upper limit is equal to the number of the remaining resource blocks over the number of secondary users. This will prevent the secondary user agent who enters the scheduling loop first from getting all the remaining resource blocks.

The secondary user agent will start sensing the remaining resource blocks that are left after scheduling the primary users. If there are remaining resource blocks, the agent will take an action from the secondary user actions set. But, if there aren't any remaining resource blocks, the agent will not take any action.

The secondary user agent deploys two types of decision policies in choosing an action. It will either explore or exploit, and this depends on the value of the exploration parameter $e$.

In the case of exploration, the agent will make a random choice in taking an action from the secondary user actions set. In the case of exploitation, the agent will make a greedy choice of what action to take from the secondary user actions set. The goal of the greedy choice is to take the action that will yield the highest reward. In order for the agent to do so, it will look through its Q-table.

**Rewards**

After executing an action, the secondary user agent will calculate the reward. The reward function for each secondary user agent is chosen to optimize the secondary user's average throughput. In this work, it was assigned to be the actual user's average throughput.

**Q-Tables**

After calculating the reward, the secondary user agent will associate this reward with the action that resulted in this reward. Then the secondary user agent will build or update its own Q-table. The agent builds its Q-table, by creating new entries as a result of exploring new actions. And the agent updates these entries as a result of choosing a pre-existing entries in the process of exploitation. The agent does these operations, according to the following formula:

$$Q(s_t, a^k) \leftarrow (1 - \alpha)\, Q(s_t, a^k) + \alpha[\, r_{t+1}(s_t, a^k) + \gamma V(s_{t+1})\, ] \qquad (7)$$

Where $V(s_{t+1})$ is determined by the policy of choosing an action at time $t$, and $a^k$ represent the action taken by secondary user $k$.

# 4. PROPOSED SOLUTION 2 – COMPETITIVE SCHEDULING ALGORITHM

The second scheduling algorithm that is proposed and implemented in Matlab is the Competitive scheduling algorithm, Algorithm 2. In this algorithm the scheduling process is also divided into two stages; the first scheduling stage is performed for the primary users, and the second one, which comes after is performed for the secondary users. It uses the competitive Q-learning approach that is based on the work proposed by [25] for both types of users, the primary and secondary users.

---

**Algorithm 2** Competitive Scheduling Algorithm

---

**Input**: no. PUs, no. SUs, no. RBs, no. epoch, exploration parameter $e$
**Output**: PUs Q-tables, SUs Q-tables
Initialize all parameters: Q-tables, PU Actions set $A_{PU}$ , SU Actions set $A_{SU}$ , state of RBs set $S$
**for** $t = 1$ to no. epoch **do**
  **for** $i = 1$ to N, where N = no. PUs **do**
    **if** no. available RBs > 0 **then**
      **if** rand < = $e$ **then**
        PU agent $i$ will explore by taking an action $a^i$ randomly from $A_{PU}$
      **else**
        PU agent $i$ will exploit by taking a greedy action $a^i$ from $A_{PU}$
      **end if**
      PU agent $i$ will calculate the reward of executing action $a^i$
      PU agent $i$ will update its Q-table based on the Q-Learning formula:
      $Q(s_t, a^i) \leftarrow (1 - \alpha) \, Q(s_t, a^i) + \alpha[(r_{t+1} \, (s_t, a^i) + \gamma \, V(s_{t+1})]$
    **end if**
    $A_{-i} = A_{PU} - \{a^i\}$
  **end for**
  **for** $k = 1$ to no. SUs **do**
    **if** no. remaining RBs in the sub-state $s_t > 0$ **then**
      **if** rand < = $e$ **then**
        SU agent $k$ will explore by taking an action $a^k$ randomly from $A_{SU}$
      **else**
        SU agent $k$ will exploit by taking a greedy action $a^k$ from $A_{SU}$
      **end if**
      SU agent $k$ will calculate the reward of executing action $a^k$
      SU agent $k$ will update its Q-table based on the Q-Learning formula:
      $Q(s_t, a^k) \leftarrow (1 - \alpha) \, Q(s_t, a^k) + \alpha[(r_{t+1} \, (s_t, a^k) + \gamma \, V(s_{t+1})]$
    **end if**
    $A_{-k} = A_{SU} - \{a^k\}$
  **end for**
**end for**

**State set S**

The set $S$ represent all the observed states of all the Resource Blocks (RBs) at all the running time.

$S = \{s_0 \, , s_1 \, , ..., s_{t-1} \, , s_t \, , s_{t+1} \, , ..., s_{no.epoch} \}$
$s_t = \{RB1 \, , ..., RBx\}$, where $x = no.RBs$
$s_t \subset S$, $s_t$ is a sub-set of the set $S$ that represent the states of all the available RBs at time $t$. And it is the same for all users agents, the primary users agents, and the secondary users agents. Each resource block has three states, free, or busy for a primary user, or busy for a secondary user.

**Primary users scheduling stage**

The primary scheduling stage of the Competitive scheduling algorithm does not include any cooperation between the primary user agents. On the contrary, the primary user agents compete on the available resource blocks. The primary users' scheduling loop of taking an action, then obtaining a reward based on executing this action, and then updating the Q-table, is repeated for every primary user agent. After the first primary user agent finishes its scheduling loop, the scheduler will update the primary user actions set to make sure that each agent will take a different action.

**Actions and Actions set**

The primary user action set $A_{PU}$ consists of multiple actions, each action has a different number of resource blocks with unique indexes. This will avoid collisions between primary user agents since they all have access to this actions set.

$A_{PU} = \{a_1^{PU}, ..., a_n^{PU}\}$, $a^{PU}$ is an action to be taken by the primary user agent, $n$ is the number of actions in $A_{PU}$. Each action has a different number of resource blocks with distinct indices, and they were set in a way to create an upper limit on how much resource blocks a primary user can get, in which a primary user can get almost fifth of the available resources at max.

The primary user agent deploys two types of decision policies in choosing an action. It will either explore or exploit, in which it depends on the value of the exploration parameter $e$. In the case of exploration, the agent will make a random choice in taking an action from the secondary user actions set. In the case of exploitation, the agent will make a greedy choice of what action to take from the secondary user actions set.

**Rewards**

After executing an action, the primary user agent will calculate the reward. The reward function for each primary user agent is chosen to optimize the primary user's average throughput. In this algorithm, it was assigned to be the actual user's average throughput.

**Q-tables**

After calculating the reward, the primary user agent will associate this reward with the action that resulted in this reward. Then the primary user agent will build or update its own Q-table. The agent builds its Q-table, by creating new entries as a result of exploring new actions. And the agent updates these entries as a result of choosing a pre-existing entries in the process of exploitation. The agent does these operations, according to the following formula:

$$Q(s_t, a^i) \leftarrow (1 - \alpha) \, Q(s_t, a^i) + \alpha[r_{t+1} \, (s_t, a^i) + \gamma \, V(s_{t+1})] \tag{8}$$

Where $V(s_{t+1})$ is determined by the policy of choosing an action at time $t$, and $a^i$ represent action $a$ taken by primary user $i$.

**Secondary users scheduling stage**

The secondary users scheduling stage in the Competitive scheduling algorithm are done exactly as the secondary users scheduling stage in the Collaborative scheduling algorithm.

# 5. Experiments

## 5.1. Experimental Setup

### 5.1.1. Experiments' fixed and variable parameters

The Network deployment of the experiments' setup consists of one macro-cell which serves 5 PUs and 5 SUs. According to the 3GPP standards and specifications, the recommended range, transmission power, and bandwidth of an LTE cellular network Macro-cell is a 1Km of diameter that is served with an eNodeB of 43 dB power, and has a bandwidth of 15 MHz, which means there will be a total of 75 Resource Blocks, each with a bandwidth of 0.2 MHz.

The learning rate α determines how much percentage of the new Q-value we want to update the previous Q-value, and it reflects the impact of the learning in updating the Q-table values. It is common in the literature to set α between 0.8 and 0.9, and in these experiments it was set to 0.8. The discount factor γ encourage the learning agent to seek out rewards sooner than later. For instance, if γ is 1, it will inform the learning agent that getting a high reward in the far future is as important as getting a high reward in the current time. Also, if γ is less than 1, and lower its value, the more important for the learning agent to get a high reward sooner than later. And if γ is equal to 0, the learning agent will only care about the immediate reward, and completely ignore rewards in the future. It is common in the literature to set γ to 0.9, and that's what it was set to in these experiments.

The exploration parameter $e$ was varied in 0.2, 0.5, and 0.8. The Number of Epochs was set to 100 because both algorithms were able to converge very quickly. And each experiment was run 200 times and their results were averaged.

### 5.1.2. Experimental Scenarios

Three main experiment scenarios were applied to both algorithms, in-order to compare their performances, and to find out the best exploration parameter $e$ for both algorithms. The performance measurements were based on the throughput percentages that each user acquired from the total macro-cell bandwidth, and the fairness level of sharing the spectrum among users. The key difference in both of these experiments' scenarios is the exploration and exploitation probabilities, these probabilities are determined by the exploration factor $e$. In the first experiment's scenario, it was set to $e = 0.5$. In the second experiment's scenario, it was set to $e = 0.2$. In the third experiment's scenario, it was set to $e = 0.8$. In all of these experiments' scenarios, the primary users had a full access to the total spectrum without any limitations on each user (e.g. any primary user could use 100% of the available resource blocks at any time), and the secondary users were allowed to access what is left of the available resource blocks of the spectrum after scheduling the primary users.

### 5.1.3. Experiments' Objective

The aim of these experiments is to test and compare the performance of both scheduling algorithms, and to measure how much of the spectrum each will utilize when there is 100% demand on the spectrum. The performance measurements were based on the throughput percentages that each user acquired from the total macro-cell bandwidth, and the fairness level of sharing the spectrum among users.

## 5.2. Experimental Results

The results are displayed in Figures 4, 5, 6, 7, 8, and 9. In all figures, the simulation results display the throughput percentages on the Y-axis and the Number of Epochs on the X-axis. The "PUs percentages" are five curves, in which each curve represents the percentage that each primary user obtained from the spectrum. The "SUs percentages" are five curves, in which each curve represents the percentage that each secondary user obtained from the spectrum. The Total percentages curve represents the sum of all the ten percentages.

### 5.2.1. The First Scenario Applied to both Scheduling Algorithms with an Exploration Parameter $e$ = 0.5

As regards to the use of the Collaborative scheduling algorithm, its performance results are displayed in Figure 4. As results shows, the algorithm converged very quickly to 91% utilization of the spectrum total throughput. This is because the scheduling algorithm is based on the use of the Q-learning formula, which is known to converge quickly to an optimal solution "choosing the optimal joint action of the primary users" as long as all actions are repeatedly sampled in all states and the action-values are represented discretely [26], which is what we followed in building our system model. This utilization of the spectrum consisted of the sum of all the users' percentages. However, all of this spectrum utilization resulted from what the primary users acquired, and this was because they had a higher priority of being scheduled over the secondary users. The distribution of the available resources followed a fair approach in which each primary user of the five had almost 18% of the spectrum, and this is because the scheduling process was based on actions that were influenced by the reward that is based on the Jain's fairness index. This helps the primary user agents to make a better joint action of how the available resources should be shared. And eventually lead to a fair distribution of the available resources and higher utilisation of the spectrum. The remaining spectrum resources "the 9% of the spectrum" are to be competed off by the secondary users. Since this is a low percentage, it wasn't enough for the secondary users to acquire a significant amount that could be displayed in the figure.

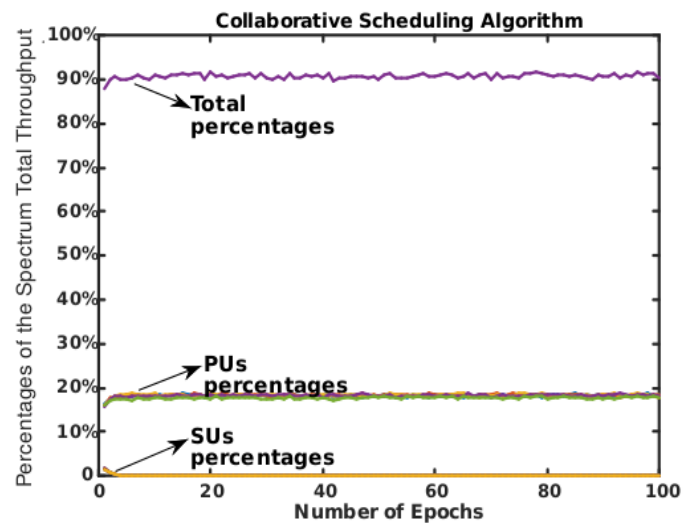

Figure 4. Percentages of the total system throughput usage while using the Collaborative scheduling algorithm with an exploration parameter of e = 0.5

As regards the use of the Competitive scheduling algorithm, its performance results are displayed in Figure 5. As results shows, the algorithm also converged very quickly to 88\% utilization of the spectrum total throughput. This utilization of the spectrum consisted of the sum of all the users' percentages. The PUs percentages resulted in 75% of the spectrum, each primary user obtained 15% of the spectrum. The distribution of the resources followed a fair approach without the use of Jain's fairness as the function reward because there was an upper limit on how much resources each primary user can get by controlling the number of resources that the actions in the actions sets can provide, in which each primary user can get almost fifth of the spectrum resources at max. About the secondary users, they got 25% of the spectrum to compete over. And they were able to obtain 15% of the spectrum, each secondary user obtained 3% of the spectrum. The fairway of distributing the remaining resources was also due to forcing a limit on how much each secondary user can get at max, in which each one of them can get almost fifth of the remaining resources at max.

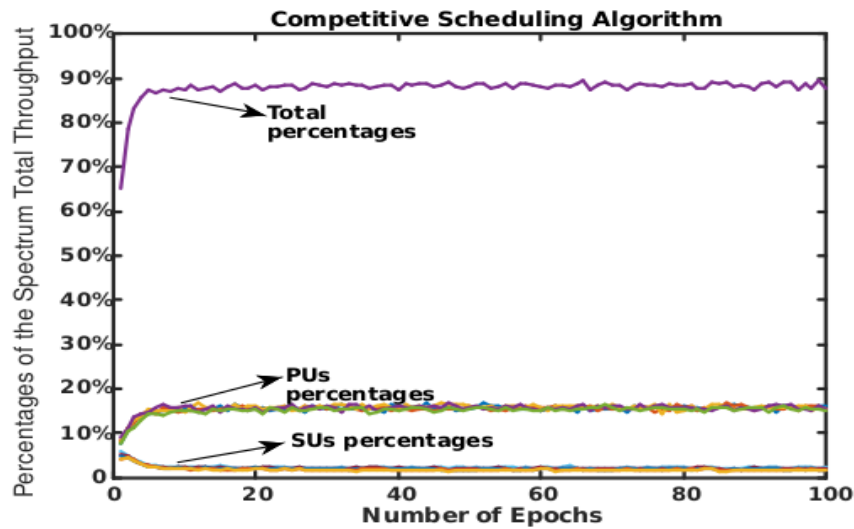

Figure 5. Percentages of the total system throughput usage while using the Competitive scheduling algorithm with an exploration parameter of e = 0.5

### 5.2.2. The Second and Third Scenario Applied to the Collaborative Algorithm with $e = 0.2$ and $e = 0.8$

As regards the use of the Collaborative scheduling algorithm, its performance results are displayed in Figure 6 and Figure 7. In both of the experiments' setups, the algorithm converged very quickly to different utilisation percentages. In the first experiment's set-up when the exploration parameter was $e = 0.2$, the algorithm converged to a 85% utilization of the spectrum total throughput. This utilization of the spectrum consisted of the sum of all the users' percentages. However, all of this utilisation resulted from what the primary users could obtain, in which it was distributed among them in a fair share. Each primary users obtained 17% of the spectrum. In the second experiment's set-up when the exploration parameter was $e = 0.8$, the algorithm converged to a 93% utilization of the spectrum total throughput. Also in this experiment's set-up, all of this utilisation resulted from what the primary users could obtain, in which it was distributed among them in a fair share. Each primary users obtained 19% of the spectrum. This total 8% increase of the spectrum is due to increasing the exploration parameter. The more exploration the user agent does, the more experience and knowledge it obtains. This will help user agents to make a better joint action of how the available resources should be shared. The choice of what joint action to take is based on the obtained reward that is calculated according to Jain's fairness index. This will eventually lead to higher utilization of the spectrum total throughput.

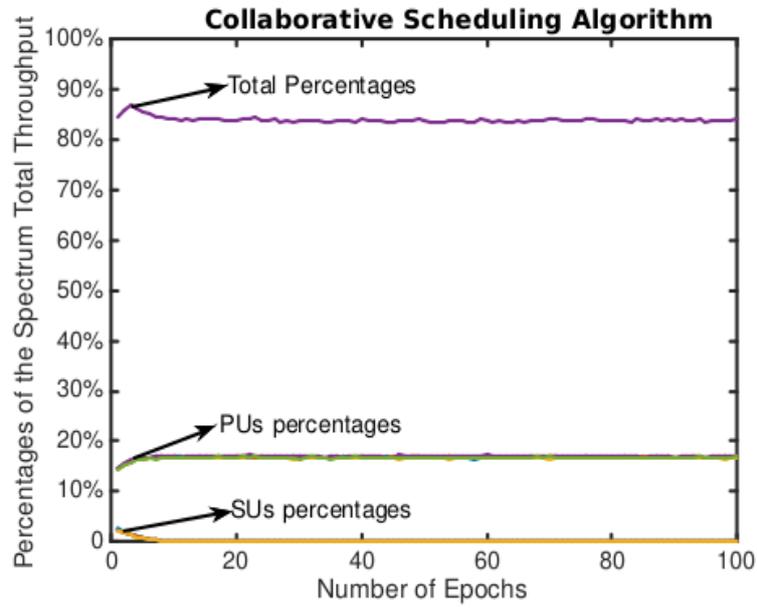

Figure 6. Percentages of the total system throughput usage while using the Collaborative scheduling algorithm with an exploration parameter of e = 0.2

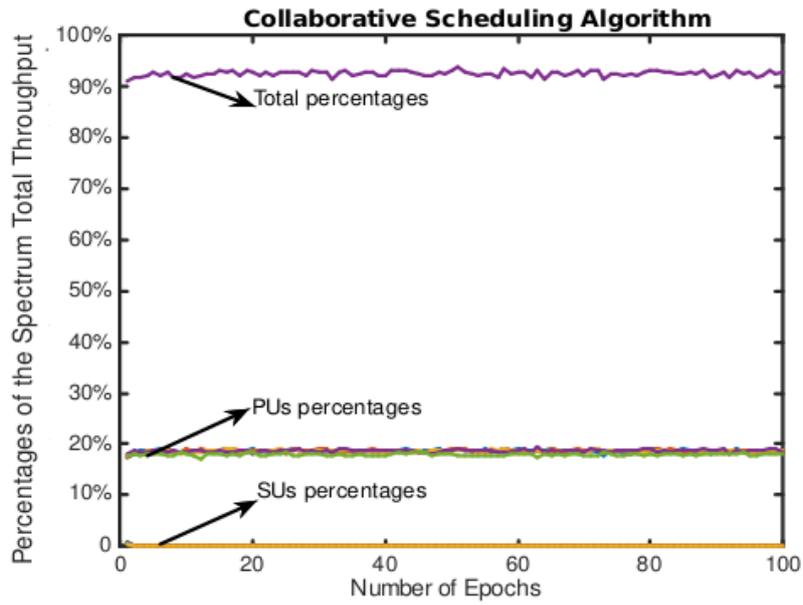

Figure 7. Percentages of the total system throughput usage while using the Collaborative scheduling algorithm with an exploration parameter of e = 0.8

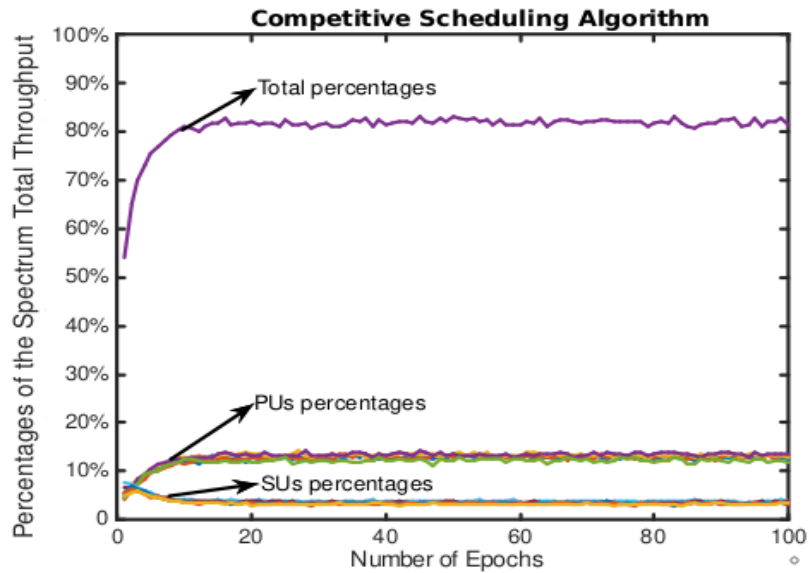

Figure 8. Percentages of the total system throughput usage while using the Competitive scheduling algorithm with an exploration parameter of e = 0.2

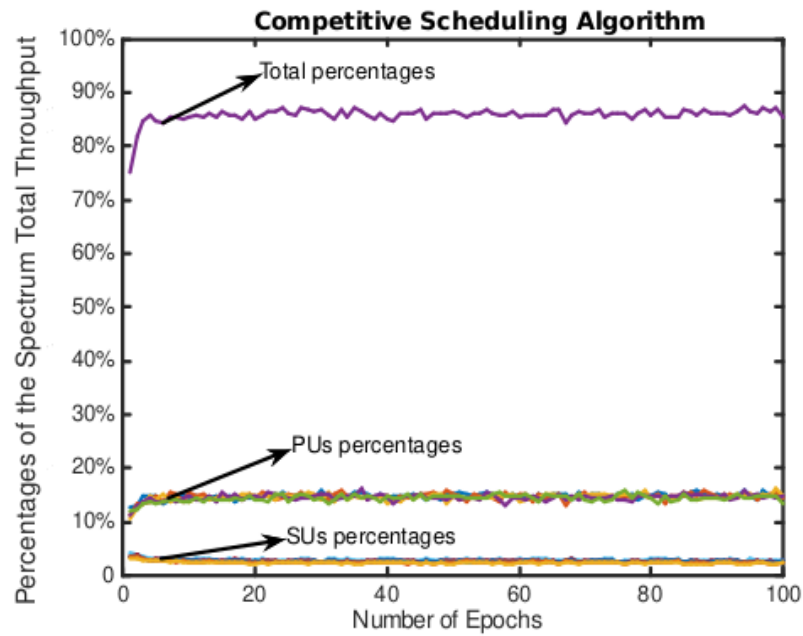

Figure 9. Percentages of the total system throughput usage while using the Competitive scheduling algorithm with an exploration parameter of e = 0.8

### 5.2.3. The Second and Third Scenario Applied to the Competitive Algorithm with *e* = 0.2 and *e* = 0.8

As regards the use of the Competitive scheduling algorithm, its performance results are displayed in Figure 8 and Figure 9. In both of the experiments' setups, the algorithm also converged very quickly and to different utilization percentages. According to Figure 8, in the first experiment's set-up when the exploration parameter was *e* = 0.2, the algorithm converged to 82% utilization of the spectrum. The PUs percentages resulted in 60% of the spectrum, 12% for each primary user.

This meant that the secondary users had more radio resources to compete over. This allowed the secondary users to obtain 22% of the spectrum total throughput, 4% for each secondary user. According to Figure 9, in the second experiment's set-up when the exploration parameter was $e = 0.8$, the algorithm converged to 85% utilization of the spectrum. The PUs percentages resulted in a 70% of the spectrum, 14% for each primary user. The SUs percentages resulted in a 15% of the spectrum total throughput, 3% for each secondary user.

# 6. CONCLUSION

In this paper, we proposed, implemented, and tested two novel scheduling algorithms. The Collaborative scheduling algorithm, and the Competitive scheduling algorithm. These algorithms schedule two types of users; the primary users, which represent the licensed subscribers, and the secondary users, which represent the unlicensed subscribers. The implementation and testing were done using Matlab. Testing the performance measurements was based on the throughput percentages that each user acquired from the total macro-cell bandwidth and the fairness level of sharing the spectrum among users. Experimental results showed that both scheduling algorithms converged to almost 90% utilization of the spectrum. However, the Collaborative scheduling algorithm provided all this utilization of the spectrum to the primary users, in which they pay for their service. In terms of distributing the resources in fair shares among users, both algorithms provided an equal degree of fairness. However, they differed in their mechanism of doing so. The Collaborative scheduling algorithm forced the fairness by using the Jains fairness index as the reward calculation for the joint action. The Competitive scheduling algorithm forced fairness among the users by creating an upper limit on how much each user can get by controlling the amount of resources that the actions in the actions sets can provide. In conclusion, it is recommended to use the Collaborative scheduling algorithm due to the high utilization of the spectrum which it can provide to the primary users, and due to the high fairness degree of distributing the resources among the primary users without the need of using an upper limit on how much each user can get. Also, our results show that the spectrum band could be utilized by deploying efficient packet scheduling algorithms for licensed users, and can be further utilized by allowing unlicensed users to be scheduled on spectrum holes whenever they occur.

## CONFLICTS OF INTEREST

The authors declare no conflict of interest.